\documentclass[12pt]{iopart}

\usepackage{bm}

\usepackage{bbm}

\usepackage{iopams}
\usepackage{graphicx,float}
\usepackage{color}
\usepackage{psfrag}
\newcommand{\be}{\begin{equation}}
\newcommand{\ee}{\end{equation}}
\newcommand{\bea}{\begin{eqnarray}}
\newcommand{\eea}{\end{eqnarray}}

\eqnobysec
\newcommand{\mc}{\mathcal}

\begin{document}

\title{Exactly conserved quasilocal operators for the XXZ spin chain}
\author{R. G. Pereira}
\address{Instituto de F\'{i}sica de S\~ao Carlos, Universidade de S\~ao Paulo, C.P. 369, S\~ao Carlos, SP,  13560-970, Brazil}
\author{V. Pasquier}
\address{Institut de Physique Th\'eorique, DSM, CEA, URA2306 CNRS, Saclay, F-91191 Gif-sur-Yvette, France}
\author{J. Sirker}
\address{Department of Physics and Research Center OPTIMAS, Technical University Kaiserslautern, D-67663 Kaiserslautern, Germany}
\address{Department of Physics and Astronomy, University of Manitoba, Winnipeg, Manitoba, Canada R3T 2N2}
\author{I. Affleck}
\address{Department of Physics and Astronomy, University of British Columbia, Vancouver, B.C., Canada, V6T 1Z1}

\date{\today}
\begin{abstract}
We extend T. Prosen's construction of quasilocal conserved quantities for the XXZ model [Phys. Rev. Lett. {\bf 106}, 217206 (2011)] to the case of periodic boundary conditions.  These quasilocal operators stem from  a two-parameter transfer matrix which employs a highest-weight representation of the quantum group algebra inherent in the Yang-Baxter algebra.  In contrast with the open chain, where the conservation law is weakly violated by boundary terms,   the quasilocal operators in the periodic chain exactly commute  with the Hamiltonian and other local conserved quantities.

\end{abstract}

\maketitle

\section{Introduction} 

Although a precise definition of quantum integrability is yet to be formulated,  the common view is that quantum integrable models are characterized by  a macroscopic number of local conserved quantities \cite{sutherland}. The most familiar examples are Bethe ansatz solvable models in which a family of commuting operators can be derived by taking logarithmic derivatives of the transfer matrix with respect to a spectral parameter \cite{korepinbook, chowdhury}. Recently the physical consequences  of nontrivial conservation laws have become  relevant  for nonequilibrium dynamics of cold atomic gases confined to one-dimensional geometries  \cite{kinoshita, trotzky}. It is generally believed that the long time average of local observables in integrable systems after a quantum quench is described  by a generalized Gibbs ensemble (GGE) which incorporates   conserved quantities besides the Hamiltonian \cite{rigol, polkovnikov}.  However, it is not clear what are all the conserved quantities that need to be included in the density matrix of the GGE. While the set of all projection operators onto   eigenstates of the Hamiltonian --- whose number increases exponentially with system size --- is definitely more than   necessary  to describe the equilibration of local observables,   the family of local conserved quantities derived from the transfer matrix --- whose number scales only linearly with system size ---  may not be sufficient for this purpose \cite{sirkerquench,fagotti1}. In fact, recent results indicate that the GGE that includes only local conserved quantities fails to  describe the steady state after a quench in the XXZ model \cite{wouters, mierzejewski,pozsgay}. GGE  expectation values for the XXZ model  had been computed before in \cite{fagotti}, where  the small deviations from numerics were attributed to  large relaxation times. New integrals of motion have also been constructed to explain why many-body localized systems (which are not integrable in the usual sense) do not thermalize \cite{serbyn,huse}.

The question of additional conserved quantities beyond the local ones usually  associated with integrability has become even more pertinent since  the discovery of a conserved \emph{quasilocal} operator for the XXZ model \cite{prosen1}.  Using a matrix product ansatz, Prosen  constructed a non-Hermitean operator that commutes with the Hamiltonian  of an open XXZ chain up to boundary terms.  The operator $Z$ in Ref. \cite{prosen1} is not  local in the usual sense because it cannot be written  in the form $Q_n=\sum_{j=1}^{N} q_j^n$, where $q_j^n$ is a local density acting on sites $j+1,\dots,j+n$ with $n$ finite. Nevertheless, it is quasilocal in the sense that the  operator norm defined at infinite temperature as  $\langle Z^\dagger Z\rangle=2^{-N}\textrm{Tr}\{Z^\dagger Z\}$  grows linearly with system size, as it does for any local operator. This quasilocal operator  cannot be written as a linear combination of the local conserved quantities obtained from the transfer matrix because it has different symmetry properties. In particular, its imaginary part changes sign under   spin inversion  $\sigma_j^z\to -\sigma_j^z,\sigma_j^\pm\to \sigma_j^\mp$, whereas the local conserved quantities are all invariant under the same transformation.  This symmetry is important because it implies that, unlike  the local conserved quantities, the quasilocal operator has an overlap with   the spin current operator and provides a nonzero Mazur   bound \cite{mazur,zotos} for the spin Drude weight at high temperatures \cite{narozhny,zotosBA,alvarez,brenig,benz,sirker, karrasch,heidrich}. This result  establishes ballistic spin transport in the critical phase of  the XXZ model at zero magnetic field, except at the Heisenberg point, where the Mazur bound vanishes \cite{prosen1, prosen2, znidaric}.

In Prosen's original construction \cite{prosen1}, the conservation of the quasilocal operator followed from a set of cubic algebraic relations that the matrices in the ansatz had to satisfy. The conservation law in the open chain is broken by boundary terms, but it was argued that, due to Lieb-Robinson bounds,  the boundary terms do not affect bulk correlators in the thermodynamic limit \cite{prosenbound}. Initially   the new conserved operator seemed unrelated to the integrability of the XXZ model. However,  it was soon realized that the cubic algebraic relations can be reduced to the quadratic quantum group algebra U$_q$[SU(2)] \cite{karevski}. The latter arises naturally  in the quantum inverse scattering method,  where it is convenient to view the XXZ model as the integrable $q$-deformation of the Heisenberg model \cite{jimbo, nepomechie, pasquier,faddeev}. More recently,  Prosen and Ilievski \cite{prosenilievski}  made an explicit connection with integrability using a highest-weight Yang-Baxter transfer operator to derive   a continuous family of quasilocal operators $Z(\varphi)$ labeled by a complex parameter $\varphi$.   This family contain the previously found operator as the particular choice $\varphi=\pi/2$. 

In this work we provide an alternative derivation of quasilocal operators that works for the periodic chain.  In this case the conservation law is not spoiled by boundary terms.\footnote{After this work had been submitted, a new paper by Prosen \cite{prosennew} appeared on the arXiv which also discusses the exact conservation of quasilocal operators for the periodic chain.}
 The paper is organized as follows. In section \ref{sectionlocal}, we review the derivation of the local conserved quantities of the XXZ model within the standard approach of taking logarithmic derivatives of the transfer matrix. In section \ref{sec:quasilocal}, we construct a  family  of two-parameter conserved quantities using an auxiliary transfer matrix with a highest-weight representation in the auxiliary space. The expansion of the conserved quantities about special values of the representation parameter, along with a discussion of the conditions that lead to quasilocality,  is presented  in section \ref{seccomputenorm}. Section \ref{sec:nonlocal} makes the point that    quasilocal operators are obtained only at  first order in the expansion about the special representation, as higher order operators are strictly nonlocal. Section \ref{sec:mazur} contains the calculation of the Mazur bound for the spin Drude weight using a single quasilocal conserved quantity. In section \ref{sec:family}, we discuss the  family of quasilocal operators obtained by varying the spectral parameter continuously. Finally, section \ref{sec:concl} presents the conclusions.

\section{Local conserved quantities\label{sectionlocal}}
Our goal is to derive generating functions of operators that commute with  the XXZ Hamiltonian \cite{korepinbook} \be
H=\sum_{j=1}^{N}\left(\sigma^x_j\sigma^x_{j+1}+\sigma^y_j\sigma^y_{j+1}+\Delta \sigma^z_j\sigma^z_{j+1}\right),\label{XXZ}
\ee
where  $\sigma^{x,y,z}$ denote the standard Pauli matrices. The Hamiltonian acts on the tensor product of vector spaces $V^{\otimes N}\equiv V_1\otimes V_2\otimes \dots\otimes V_N$, where $V_j=\mathbb C^2$  is the quantum space of the spin on site $j$. 

In the general scheme of the quantum inverse scattering method, one starts by introducing an $R$ matrix that depends on a complex spectral parameter $z$ and satisfies the Yang-Baxter equation  \cite{korepinbook,chowdhury}\be
R_{12}(zw^{-1})R_{1\mc Q}(z)R_{2\mc Q}(w)=R_{2\mc Q}(w)R_{1\mc Q}(z)R_{12}(zw^{-1}).\label{YB1}
\ee
Here $R_{12}(z)$ acts nontrivially on $V_1\otimes V_2$ and as the identity on a third,  auxiliary space $\mc Q$ with  dimension $d_{\mc Q}$, to be specified below.   For the XXZ model (or six-vertex model), we can write $R_{12}(z)=R(z)\otimes \mathbbm 1$ with the $R$ matrix \be
R(z)=\left(\begin{array}{cccc}a&&& \\
&cz^{-1}&b&\\
&b&cz&\\
&&&a
\end{array}\right),\label{Rmatrixsixvertex}
\ee
with $a=zq-z^{-1}q^{-1}$, $b=z-z^{-1}$, $c=q-q^{-1}$. The parameter $q$ is related to the anisotropy  $\Delta$ in   Eq. (\ref{XXZ}) by $\Delta=(q+q^{-1})/2$. We use the following  notation for matrices that act on the tensor product of two spaces \cite{chowdhury} :\be
R=R^{ij}_{kl}\,e^{V_1}_{ij}\otimes e^{V_2}_{kl}.
\ee
Here $e^{V_1}_{ij}$ are matrices acting on $V_1$ defined by $e^{V_1}_{ij}=\hat e_i\otimes \hat e_j$, where the set of vectors $\{\hat e_i\}$ forms an orthonormal  basis of $V_1$. In the example of the tensor product of   two-dimensional spaces ($i,j=1,2$), the $R$ matrix in Eq. (\ref{Rmatrixsixvertex}) is written in the form\be
R(z)=\left(\begin{array}{cccc}R^{11}_{11}&R^{11}_{12}&R^{12}_{11}&R^{12}_{12} \\
R^{11}_{21}&R^{11}_{22}&R^{12}_{21}&R^{12}_{22}\\
R^{21}_{11}&R^{21}_{12}&R^{22}_{11}&R^{22}_{12}\\
R^{21}_{21}&R^{21}_{22}&R^{22}_{21}&R^{22}_{22}
\end{array}\right).
\ee
The product of $e^{V_1}_{ij}$ matrices has the property  $e^{V_1}_{ij}e^{V_1}_{kl}=\delta_{jk}e^{V_1}_{il}$. Using this property we can write down  each element of the Yang-Baxter equation (\ref{YB1}) corresponding to $e^{V_1}_{ij}\otimes e^{V_2}_{kl}\otimes e^{\mc Q}_{mn}$, with $i,j,k,l=1,2$ and $m,n=1,\dots, d_{\mc Q}$ :\bea
&&\sum_{a,b=1}^2\sum_{c=1}^{d_{\mc Q}}[R_{12}(zw^{-1})]^{ia}_{kb}[R_{1\mc Q}(z)]^{aj}_{mc}[R_{2\mc Q}(w)]^{bl}_{cn}\nonumber\\
&=&\sum_{a,b=1}^2\sum_{c=1}^{d_{\mc Q}}[R_{2\mc Q}(w)]^{kb}_{mc}[R_{1\mc Q}(z)]^{ia}_{cn}[R_{12}(zw^{-1})]^{aj}_{bl}.
\eea

The Lax operator associated with a given  site $j$ can be introduced as an $R$ matrix that acts on $V_j\otimes \mc Q$\be
L_j(z)=R_{j\mc Q}(z).
\ee
Then Eq. (\ref{YB1}) implies the quadratic relation for Lax operators involving the $R$ matrix in Eq.  (\ref{Rmatrixsixvertex})
\be
R_{12}(zw^{-1})L_1(z)   L_2(w)=L_2(w)  L_1(z)R_{12}(zw^{-1}).\label{YB}
\ee
It is convenient to  express the solutions of Eq. (\ref{YB}) for an arbitrary auxiliary space in terms of operators $K,S^+,S^-$:   \bea
L_j(z)&=&\frac12\left[\frac{(z-z^{-1})}2\,\mathbbm 1_j\otimes \mathbbm (K+K^{-1})+\frac{(z+z^{-1})}2\,\sigma_j^z\otimes (K-K^{-1})\nonumber\right.\\
&&\left.+(q-q^{-1})(z\sigma_j^+\otimes S^-+z^{-1}\sigma_j^-\otimes S^+)\right].\label{Laxop}
\eea
Written as a matrix  in $V_j$ (with entries that act on $\mc Q$), the Lax operator is\be
L_j(z)=\frac12\left(\begin{array}{cc}zK-z^{-1}K^{-1}&z(q-q^{-1})S^-\\
z^{-1}(q-q^{-1})S^+&zK^{-1}-z^{-1}K\end{array}\right).\label{Laxmatrix}
\ee
The Yang-Baxter equation (\ref{YB}) is then satisfied provided  that the operators  $S^\pm $, $K$ acting on $\mc Q$  obey the quantum group algebra U$_q$[SU(2)]\cite{faddeev}\bea
KS^+&=&qS^+K,\label{quantum1}\\
KS^-&=&q^{-1}S^-K,\label{quantum2}\\
\left[S^+,S^-\right]&=&\frac{K^2-K^{-2}}{q-q^{-1}}.\label{quantum3}
\eea

Choosing $\mc Q=\mathbb C^2$, we can use the spin-1/2 representation  \be
K=q^{\tau^z/2},\qquad S^\pm=\tau^\pm,\label{2Drep}\ee
where $\tau^{x,y,z}$ are Pauli matrices in the auxiliary space. The monodromy matrix  for $N$ spins  (acting on $V_1\otimes V_2\otimes \dots V_N\otimes \mc Q$) is defined as \be
T_{\mc Q}(z)=L_N(z)L_{N-1}(z)\dots L_1(z).\label{monodromyQ}
\ee
The transfer matrix is \be
t_{\mc Q}(z)=\textrm{tr}_{\mc Q}\{T_{\mc Q}(z)\},\label{transfermatrix}
\ee
where $\textrm{tr}_{\mc Q}$ denotes the trace over the auxiliary space $\mc Q$. It can be shown  \cite{korepinbook,chowdhury} that  the transfer matrix forms a one-parameter family of commuting   operators in $V_1\otimes V_2\otimes \dots\otimes V_N$:\be
[t_{\mc Q}(z),t_{\mc Q}(w)]=0,\qquad \forall\, z,w\in\mathbb C.
\ee 

The local conserved quantities $Q_n$ are  given by \cite{korepinbook}\be
Q_{n+1}=\left.\frac{d^n}{dz^n}\ln t_{\mc Q}(z)\right|_{z=1},\qquad n\geq 1.\label{localQsfromt}
\ee
The operator $Q_2$ is proportional to  the XXZ Hamiltonian in Eq. (\ref{XXZ}). The first nontrivial conserved quantity, $Q_3$, coincides with the energy current operator \cite{klumper}. In general, each $Q_n$ can be written as a sum of operators  that act on $n$ neighbouring spins and is therefore local.

Consider   the spin inversion transformation $\mc C$ defined in the quantum space as $\mc C^{-1}\sigma_j^z\mc C=-\sigma_j^z$, $\mc C^{-1}\sigma_j^\pm \mc C=  \sigma_j^\mp$, $\forall j$. The Lax operator   in Eq. (\ref{Laxop})  transforms  as\bea
\tilde L_j(z)&= &\mc C^{-1}L_j(z) \mc C\nonumber\\
&=&\frac12\left[\frac{(z-z^{-1})}2\,\mathbbm 1_j\otimes \mathbbm (K+K^{-1})-\frac{(z+z^{-1})}2\,\sigma_j^z\otimes (K-K^{-1})\right.\nonumber\\
&&\left.+(q-q^{-1})(z\sigma_j^-\otimes S^-+z^{-1}\sigma_j^+\otimes S^+)\right].
\eea
In the case of the spin-1/2 representation in Eq. (\ref{2Drep}), we can show that the transfer matrix is invariant under spin inversion using the following similarity  transformation  that acts on   $\mc Q$:\be
W(z)=[W(z)]^{-1}=\left(\begin{array}{cc}0 &z\\
z^{-1}&0\end{array}\right).
\ee
This transformation is such that \bea
\left[W(z)\right]^{-1} q^{\tau^z/2} W(z)&=&q^{-\tau^z/2},\\
\left[W(z)\right]^{-1} \tau^+ W(z)&=&z^{2}\tau^-,\\
 \left[W(z)\right]^{-1} \tau^-W(z)&=&z^{-2}\tau^+.
\eea
It follows that \be
[W(z)]^{-1}\tilde L_j(z)W(z)=L_j(z).\label{similarity}
\ee
As a result,\bea
\tilde t_{\mc Q}(z)&=&\mc C^{-1} t_{\mc Q}(z)\mc C\nonumber\\
&=&\textrm{tr}_{\mc Q}\{\tilde L_N(z)\tilde L_{N-1}(z)\dots \tilde L_1(z)\}\nonumber\\
&=&\textrm{tr}_{\mc Q}\{W^{-1}\tilde L_NWW^{-1}\tilde L_{N-1}W\dots W^{-1}\tilde L_1W\}\nonumber\\
&=&\textrm{tr}_{\mc Q}\{ L_N(z)  L_{N-1}(z) \dots   L_1(z)\}\nonumber\\
&=&t_{\mc Q}(z).\label{svinvariance}
\eea
Since Eq. (\ref{svinvariance}) is verified for all $z\neq0,\infty$, we conclude that all the $Q_n$'s derived by expanding   $t_{\mc Q}(z)$ about $z=1$  are invariant under spin inversion. For instance, this is clearly the case for the XXZ Hamiltonian at zero magnetic field in Eq. (\ref{XXZ}).

\section{Conserved quantities from two-parameter transfer matrix\label{sec:quasilocal}}

The idea to obtain  a generating function of conserved quantities which are \emph{not} invariant under spin inversion is to introduce an auxiliary transfer matrix that commutes with $t_{\mc Q}(z)$ but employs  a different representation of the quantum group algebra. Let us consider an auxiliary space $\mc A $ with   dimension $d_{\mc A}$. We denote the Lax operator defined in $V_j\otimes \mc A$ by \be
\mc L_j(z)=R_{j\mc A}(z).\label{laxinA}
\ee
By analogy with Eq. (\ref{monodromyQ}), we can define the corresponding monodromy matrix\be
T_{\mc A}(z)=\mc L_N(z)\mc L_{N-1}(z)\dots \mc L_1(z),
\ee
as well as   the auxiliary transfer matrix \be
t_{\mc A}(z)=\textrm{tr}_{\mc A}\{T_{\mc A}(z)\}.\label{taux}
\ee
We then apply the ``train argument''\cite{faddeev} for the Yang-Baxter equation with an $R$ matrix in $\mc Q\otimes \mc A$ as follows:\bea
T_{\mc Q}(z)T_{\mc A}(w)R_{\mc Q\mc A}(w/z)&=& R_{N\mc Q}(z)\dots   R_{1\mc Q}(z)  R_{N\mc A}(w)\dots   R_{1\mc A}(w)R_{\mc Q\mc A}(w/z)\nonumber\\
&=&R_{N\mc Q}\dots R_{2\mc Q}    R_{N\mc A}\dots  R_{2\mc A}R_{1\mc Q}R_{1\mc A}R_{\mc Q\mc A}\nonumber\\
&=&R_{N\mc Q}\dots R_{2\mc Q}    R_{N\mc A}\dots  R_{2\mc A}R_{\mc Q\mc A}R_{1\mc A}R_{1\mc Q}\nonumber\\
&=&R_{\mc Q\mc A}R_{N\mc A}\dots  R_{1\mc A}R_{N\mc Q}\dots R_{1\mc Q}   \nonumber\\
&=& R_{\mc Q\mc A}(w/z) T_{\mc A}(w)T_{\mc Q}(z).\label{trainQA}
\eea
Taking the trace of Eq. (\ref{trainQA}) over $\mc Q$ and $\mc A$, we obtain 
\be
[\textrm{tr}_{\mc Q}\{T_{\mc Q}(z)\},\textrm{tr}_{\mc A}\{T_{\mc A}(w)\}]=0,
\ee
thus\be
[t_{\mc Q}(z),t_{\mc A}(w)]=0\qquad \forall\, z,w\in\mathbb C.
\ee
Therefore, since the XXZ Hamiltonian is among the operators generated by $t_{\mc Q}(z)$, we can use $t_{\mc A}(z)$ as a generating function of conserved quantities.

We shall work with  the highest weight representation of U$_q$[SU(2)]:\bea
K|r\rangle&=& uq^{r}|r\rangle,\label{Khw}\\
S^{+}|r\rangle&=&-a_r|r+1\rangle,\label{Sphw} \\
S^{-}|r\rangle&=&b_r|r-1\rangle,\label{Smhw} 
\eea
where $u\in \mathbb C$ is    arbitrary. The index $r$ can be interpreted as positions in a lattice in the auxiliary space, and the operators $S^+$ and $S^-$ perform   hopping  between nearest-neighbour sites.  Eq. (\ref{quantum3}) imposes the relation \be
a_rb_{r+1}-a_{r-1}b_r=\frac{u^2q^{2r}-u^{-2}q^{-2r}}{q-q^{-1}},
\ee
which is satisfied by the choice\bea
a_r&=&v\,\frac{u^2q^r- u^{-2}q^{-r}}{q-q^{-1}},\\
b_r&=&v^{-1}\, \frac{q^r-q^{-r}}{q-q^{-1}},
\eea
where $v$ is another arbitrary parameter which we set to 1 hereafter. In this representation  the Casimir operator is a function of the parameter $u$:\be
C=(q-q^{-1})^2S^+S^-+q^{-1}K^2+qK^{-2}=u^2q^{-1}+u^{-2}q.
\ee

The dimension of the auxiliary space depends on the value of $\Delta=(q+q^{-1})/2$.
When $q$ is a root of unity, {\it i.e.} $q=e^{i\lambda}$ with $\lambda=l\pi/m$ and  $l,m\in \mathbb Z$ coprimes, we have $b_0=b_m=0$. In these cases we can restrict the auxiliary space index $r$ to  $0\leq r\leq m-1$ and  the  representation has finite dimension $d_{\mc A}=m$.  Notice that for $q=e^{i\pi l/m}$ we have    $\Delta=\cos(\pi l/m)$, hence $|\Delta|\leq 1$, which corresponds to the gapless phase of the XXZ model.

The matrices in Eqs. (\ref{Khw}), (\ref{Sphw}) and  (\ref{Smhw}) are functions of  the complex parameter $u$. The Lax operator defined in Eq. (\ref{laxinA}) is a function of both $u$ and the spectral parameter $z$. Similarly to Eq. (\ref{Laxop}), we can write \bea
\mc L_j(z,u)&=&i[\mathbbm 1_j\otimes A_0(z,u)+\sigma_j^z\otimes A_z(z,u)\nonumber\\
&&+\sigma_j^+\otimes A_+(z,u)+\sigma_j^-\otimes A_-(z,u)],
\eea
where\bea
A_0(z,u)&=&\frac{(z-z^{-1})}{4i} \mathbbm [K(u)+K^{-1}(u)],\label{A0}\\
A_z(z,u)&=&\frac{(z+z^{-1})}{4i} \mathbbm [K(u)-K^{-1}(u)],\label{Az}\\
A_+(z,u)&=&\frac z{2i}(q-q^{-1})S^-(u),\label{Ap}\\
A_-(z,u)&=&\frac {z^{-1}}{2i}(q-q^{-1})S^+(u).\label{Am}
\eea
In this notation, the conserved quantity defined in Eq. (\ref{taux}) reads (hereafter we omit the index $\mc A$ in tr$_{\mc A}$) \be
t_{\mc A}(z,u)=i^N\sum_{\{\alpha_j\}}\textrm{tr}\{A_{\alpha_N}\dots A_{\alpha_2}A_{\alpha_1}\}\prod_{j=1}^{N}\sigma_j^{\alpha_j},\label{conservedtA}
\ee
where the sum is over all $\alpha_j\in\{0,z,+,-\}$ and we use the notation $\sigma_j^0\equiv \mathbbm 1_j$. 

The operator in Eq. (\ref{conservedtA}) is translationally invariant due to the cyclic property of the trace. On the other hand, it is not necessarily  invariant under spin reversal for general $u$. (The similarity between matrices    used in Eq. (\ref{similarity}) is not verified for arbitrary values of $u$.) Moreover, $t_{\mc A}(z,u)$ is not  invariant under parity transformation $\mc{P}$, which we can  define as the reflection about the link between sites $j=1$ and $j=N$:   $  \mc P^{-1}\sigma_j^{\alpha_j}\mc P= \sigma_{N+1-j}^{\alpha_j}$. We have\bea
\mc P^{-1}t_{\mc A}(z,u) \mc P&=&i^N\sum_{\{\alpha_j\}}\textrm{tr}\{A_{\alpha_N}\dots A_{\alpha_2}A_{\alpha_1}\}\prod_{j=1}^{N}\sigma_{N+1-j}^{\alpha_{j}}\nonumber\\
&=&i^N\sum_{\{\alpha_j\}}\textrm{tr}\{A_{\alpha_1}\dots A_{\alpha_{N-1}}A_{\alpha_N}\}\prod_{j=1}^{N}\sigma_{j}^{\alpha_{j}}\nonumber.\\
\eea
We note that  $t_{\mc A}$ in Eq. (\ref{conservedtA}) can also be written as \be
t_{\mc A}(z,u)=i^N\sum_{\{\alpha_j\}}\textrm{tr}\{A_{\alpha_1}^{t}A_{\alpha_2}^t\dots A_{\alpha_N}^t\}\prod_{j=1}^{N}\sigma_j^{\alpha_j},
\ee
where $A^t_\alpha$ denotes the transpose of $A_\alpha$. We define the two-parameter conserved quantity which is odd under parity  as\be
\mc I(z,u)=(-i)^N[\mc P^{-1}t_{\mc A}(z,u) \mc P-t_{\mc A}(z,u)].\label{conservedIzu1}
\ee
Therefore,\be
\mc I(z,u)=\sum_{\{\alpha_j\}}\textrm{tr}\{A_{\alpha_1}\dots A_{\alpha_N}-A_{\alpha_1}^{t}\dots A_{\alpha_N}^t\}\prod_{j=1}^{N}\sigma_j^{\alpha_j}.\label{conservedIzu}
\ee

For reference, let us comment on the particular cases     $\Delta=0$ and $\Delta=\pm1$. For  $\Delta=0$   the XXZ model is equivalent to free fermions via a Jordan-Wigner transformation. This point    corresponds to $m=2$, $q=i$; in this case the generators of the quantum group algebra become\be
K=u\left(\begin{array}{cc}
1&0\\
0&i
\end{array}\right),\qquad S^+=\frac{u^{-2}-u^{2}}{2i}\sigma^-,\qquad S^-=\sigma^+.
\ee
Note that, although the auxiliary space is two-dimensional $\mc A=\mathbb C^2$, the representation differs from Eq. (\ref{2Drep}) for general $u$. Only for $u=e^{-i\pi/4}$ do we recover a parity-invariant representation. On the other hand,  at the ferromagnetic SU(2) point $\Delta=-1$ ($q=-1$,  $m=1$),    the $A_\alpha$ matrices reduce to numbers and the conserved quantity  in Eq. (\ref{conservedIzu})   vanishes identically. At the antiferromagnetic SU(2) point $\Delta=1$ ($q=1$, $m\to \infty$) the operator is not identically zero but the representation becomes infinite dimensional.

\section{Quasilocal conserved quantities\label{seccomputenorm}}

Now we turn to the task of extracting quasilocal operators from $\mc I(z,u)$ in Eq. (\ref{conservedIzu}). In order to calculate the Mazur bound for the Drude weight at high temperatures \cite{zotos}, it is convenient to define the inner product between two operators $A$ and $B$  acting on $V^{\otimes N}$ based on the thermal average at infinite temperature:\be
 \langle A^\dagger B\rangle =2^{-N}\textrm{Tr}\{A^\dagger B\},\label{normQQ}
\ee
where Tr denotes the trace over the quantum  space $V^{\otimes N}$. From Eq. (\ref{normQQ}) it can be shown that the norm of $\mc I(z,u)$ reduces to 
\be
 \langle \mc I^\dagger(z,u) \mc I(z,u)\rangle=2\,\textrm{tr}_{\mc A\otimes\mc A}\{[T_1(z,u,u)]^N-[T_2(z,u,u)]^N\}.\label{normfromTs}
\ee
Here $T_1(z,u,\bar u)$ and $T_2(z,u,\bar u)$ are transfer matrices in $\mc A\otimes \mc A$ \bea
T_1(z,u,\bar u) &=& \sum_{\alpha=0,z,\pm}C_\alpha A^*_{\alpha}(z,u)\otimes A_\alpha(z,\bar u),\label{T1general}\\
T_2 (z,u,\bar u) &=& \sum_{\alpha=0,z,\pm}C_\alpha A^*_{\alpha}(z,u)\otimes A^t_\alpha(z, \bar u),\label{T2general}
\eea
where\be
C_\alpha=\frac12\textrm{Tr}\left\{\sigma^{\alpha}(\sigma^{\alpha})^\dagger\right\}.
\ee

In contrast with Prosen's construction for the open chain  \cite{prosen1},  where the norm is computed from the matrix element  between boundary states, Eq. (\ref{normfromTs}) involves the trace over the auxiliary space. The analogy with the open chain can be explored further if we notice that,  by setting the spectral parameter to be $z=i$, the matrix $A_z$ in Eq. (\ref{Az}) vanishes and the conserved quantity does not contain any $\sigma_j^z$ operators, as assumed in the original matrix product ansatz \cite{prosen1}. In the following we shall focus on the particular choice $z=i$. We return to the question of general values of $z$ in section \ref{sec:family}. For $z=i$ the nonvanishing operators  in  auxiliary space simplify to 
\bea
A_0(z=i,u)&=&\sum_{r=0}^{m-2}\frac{uq^r+u^{-1}q^{-r}}2|r\rangle\langle r|,\label{A0u}\\
A_+(z=i,u)&=&\sum_{r=0}^{m-2}\frac{q^{r+1}-q^{-r-1}}2|r\rangle\langle r+1|,\label{Apu}\\
A_-(z=i,u)&=&\sum_{r=0}^{m-2}\frac{u^2q^r-u^{-2}q^{-r}}{2}|r+1\rangle\langle r|.\label{Amu}
\eea

After fixing the value of the spectral parameter, we are still free to choose the value of $u$ in the representation of the quantum group algebra. We notice that  the condition $u^4=1$ is special because in this case $a_0=0$, then the state $|r=0\rangle$ is annihilated by $A_\pm$ and decouples from the other states. Hereafter we choose $u=1$, but the result for the other roots is equivalent. For $u=1$ the Casimir operator becomes $C=q+q^{-1}$. Interestingly, a similar kind of special representation appears in open spin chains where the quantum group is an actual symmetry commuting with the Hamiltonian \cite{pasquier}. In that case, the Casimir  for a spin-$1/2$ representation  takes the value $q^{N-1}+q^{-N+1}$ and becomes special if $q^N=-1$, {\it i.e.}, for values  of $q$ that obey a ``root of unity condition'' depending on the chain length \cite{mccoy}.

Let us then analyze the operator \be
\mc I_0\equiv \mc I(z=i,u=1).\label{Izero}\ee 
Setting $u=1$ in Eqs. (\ref{A0u}) through  (\ref{Amu}), we obtain (recall $q=e^{i\lambda}$)\bea
A_0(1)&\equiv&A_0(z=i,u=1)=\sum_{r=0}^{m-1}\cos(\lambda r)|r\rangle\langle r|,\\
A_+(1)&\equiv&A_+(z=i,u=1)=i\sum_{r=0}^{m-2}\sin[\lambda(r+1)]|r\rangle\langle r+1|,\\
A_-(1)&\equiv&A_-(z=i,u=1)=-i\sum_{r=0}^{m-2}\sin(\lambda r)|r+1\rangle\langle r|.
\eea
The transfer matrices in Eqs. (\ref{T1general}) and (\ref{T2general}) become \bea
T_1(1)&\equiv&T_1(z=i,u=1,\bar u=1)\nonumber\\
&=&\sum_{r,s=0}^{m-1}\cos(\lambda r)\cos(\lambda s)|r,s\rangle\langle r,s|\nonumber\\
&&+\frac12\sum_{r,s=0}^{m-2}\sin[\lambda (r+1)]\sin [\lambda (s+1)]|r,s\rangle\langle r+1,s+1|\nonumber\\
&&+\frac12\sum_{r,s=0}^{m-2}\sin (\lambda r)\sin(\lambda s)|r+1,s+1\rangle\langle r,s|,\\
T_2(1)&\equiv&T_2(z=i,u=1,\bar u=1)\nonumber\\
&=&\sum_{r,s=0}^{m-1} \cos(\lambda r)\cos(\lambda s)|r,s\rangle\langle r,s|\nonumber\\
&&+\frac12\sum_{r,s=0}^{m-2}\sin[\lambda (r+1)]\sin [\lambda (s+1)] |r,s+1\rangle\langle r+1,s|\nonumber\\
&& +\frac12\sum_{r,s=0}^{m-2}\sin (\lambda r)\sin(\lambda s)|r+1,s\rangle\langle r,s+1| .
\eea

The transfer matrices are block diagonal in subspaces of Kronecker states ${\{|r,(r+k)(\textrm{mod }m)\rangle \}}$ with fixed $k=0,\dots,m-1$ in the case of $T_1(1)$, or Kronecker states $\{|r,(-r+k)(\textrm{mod }m)\rangle\}$ in the case of $T_2(1)$. Since we are interested in the scaling of the operator norm in Eq. (\ref{normfromTs}) with system size $N$ as $N\to \infty$, we may  restrict ourselves  to the subspace in which the transfer matrices have their largest eigenvalue. This happens when $k=0$ for both $T_1(1)$ and $T_2(1)$. Within the $k=0$ subspace we denote $|r,\pm r\rangle \to |r\rangle $ and obtain the reduced transfer matrices
\bea
\mc T_1&=&\sum_{r=0}^{m-1}\cos^2(\lambda r)|r\rangle\langle r|+\frac12\sum_{r=0}^{m-2} \sin^2[\lambda (r+1)]|r\rangle\langle r+1|\nonumber\\
&&+\frac12\sum_{r=0}^{m-2} \sin^2 (\lambda r)|r+1\rangle\langle r|,\label{defT1red}\\
\mc T_2&=&\sum_{r=0}^{m-1}\cos^2(\lambda r)|r\rangle\langle r|-\frac12\sum_{r=0}^{m-2} \sin(\lambda r)\sin[\lambda (r+1)]\times\nonumber\\
&&\times[|r\rangle\langle r+1|+|r+1\rangle\langle r|].\label{defT2red}
\eea
It is useful to note that \bea
\mc T_1&=& \mathbbm 1-B^2+\frac12\Delta B^2,\label{T1withdeltanadB}\\
\mc T_2&=& \mathbbm 1-B^2-\frac12B\Delta B,\label{T2withdeltanadB}
\eea
where $B$ is the diagonal matrix $B=\sum_{r=0}^{m-1}\sin(r\lambda)|r\rangle\langle r|$
 and $\Delta$ is the uniform hopping matrix on an open chain with length $m$ \be
\Delta=\sum_{r=0}^{m-2}\left(|r\rangle\langle r+1|+|r+1\rangle\langle r|\right).
\ee

\begin{figure}
\begin{center}
\includegraphics*[width=.5\columnwidth]{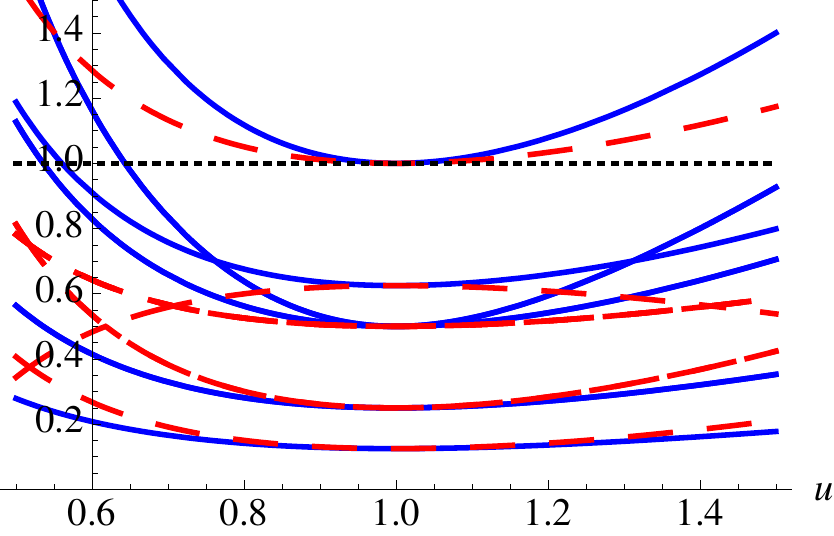}
\end{center}
\caption{Absolute value of the eigenvalues of the transfer matrices $T_1(z=i,u,u)$ (solid blue lines) and $T_2(z=i,u,u)$ (dashed red lines)  as a function of $u\in \mathbb R$ for $q=e^{i\pi/3}$.  The quasilocal conserved quantity is obtained by expanding about $u=1$, where $T_1$ and $T_2$ have the same spectrum and their largest eigenvalue is normalized to 1. \label{fig:eigenvalues}}
\end{figure}

The matrix $\mc T_2$ is symmetric, thus its eigenvalues are all real. Since $B|0\rangle=0 $, we find that $|r=0\rangle$ is an eigenvector of $\mc T_2$ with eigenvalue 1. It is easy to verify  that all the other eigenvalues are smaller than 1.\footnote{For $\lambda\in \mathbb R$, {\it i.e.} $|\Delta|\leq1$, we can show that the largest eigenvalue of $\mc T_2$  is 1 using the Gershgorin circle theorem. In the gapped Neel phase $\Delta>1$ the method used here only gives rise to nonlocal operators.} On the other hand, $\mc T_1$ is not symmetric. However, in  \ref{appsimilar} we   show that $\mc T_1$ and $\mc T_2$ are similar and have exactly the same spectrum  (see  also Fig. \ref{fig:eigenvalues}). It also follows from Eq. (\ref{T1withdeltanadB}) that $|r=0\rangle$ is the   \emph{right} eigenvector of $\mc T_1$ with eigenvalue 1. We will also need the \emph{left} eigenvector of $\mc T_1$ with eigenvalue 1. In   \ref{appeigenv} we show that the solution to the eigenvalue equation $\langle 0_L| \mc T_1= \langle 0_L|$ yields\be
\langle 0_L| =\sum_{r=0}^{m-1}(1-r/m)\langle r|.\label{zeroleft}
\ee
The left eigenvector $\langle 0_L|$ is not normalized to unity but is such that $\langle 0_L|0\rangle=1$.

When  calculating  the norm of the conserved quantity using Eq. (\ref{normfromTs}), we can use the macroscopic number of transfer matrices to project  the auxiliary space  into the eigenvectors of $\mc T_1$ or $\mc T_2$ with eigenvalue 1. 
In   \ref{appA} we show that\bea
\lim_{n\to \infty}\mc T_1^n&=&\lim_{n\to \infty}\mc T_1^{n /2}\sum_{r}|r\rangle\langle r|\mc T_1^{n/2}=|0\rangle\langle 0_L|,\label{T1projector}\\
\lim_{n\to \infty}\mc T_2^n&=&\lim_{n\to \infty}\mc T_2^{n/2} \sum_{r}|r\rangle\langle r|\mc T_2^{n/2}=|0\rangle\langle 0|.\label{T2projector}
\eea
In particular, the projection allows us to compute traces involving an arbitrary matrix $M$
\bea
\lim_{n\to \infty}\textrm{tr}\{\mc T_1^nM\}&=&\langle0_L|M|0\rangle,\\
\lim_{n\to \infty}\textrm{tr}\{\mc T_2^nM\}&=&\langle0|M|0\rangle.
\eea

It turns out that we do not get a quasilocal operator by simply setting  $z=i, u=1$. The reason is that, since $\mc T_1$ and $\mc T_2$   are related by a similarity transformation,  the operator $\mc I_0$  in Eq. (\ref{Izero}) actually has zero norm:\be
\langle \mc I_0^\dagger \mc I_0\rangle =\textrm{tr}\{\mc T_1^N-\mc T_2^N\}=0.
\ee
Nevertheless, the properties of the transfer matrices   suggest that quasilocal operators can be generated by expanding $\mc I(z=i,u)$ about $u=1$:\be
\mc I(z=i,u=1+\varepsilon)= \varepsilon \mc I_1+\mc O(\varepsilon^2).\label{expandIzu}
\ee

Let us then consider the operator\bea
\mc I_1&=&\left.\frac{\partial \mc I(z,u)}{\partial u}\right|_{z=i,u=1}\nonumber\\
&=&\sum_{\{\alpha_j\}}\textrm{tr}\left\{\frac{\partial}{\partial u}[A_{\alpha_1}(u)\dots A_{\alpha_N}(u)-A_{\alpha_1}^{t}(u)\dots A_{\alpha_N}^t(u)]\right\}\prod_{j=1}^{N}\sigma_j^{\alpha_j},\label{defI1}
\eea
with matrices $A_\alpha(u)$ given in Eqs. (\ref{A0u}), (\ref{Apu}) and (\ref{Amu}).
Using the transfer matrices in Eqs. (\ref{T1general}) and (\ref{T2general}), we can express the norm of $\mc I_1$ as follows:\bea
\langle \mc I^\dagger_1\mc I_1\rangle &=&2\left.\textrm{ tr}\left\{\frac{\partial^2}{\partial u\partial \bar u}[T_1(z,u,\bar u)]^{N}\right\}\right|_{z=i, u=\bar u=1}\nonumber\\
&&-2\left.\textrm{tr}\left\{\frac{\partial^2}{\partial u\partial \bar u}[T_2(z,u,\bar u)]^{N}\right\}\right|_{z=i, u=\bar u=1}.\label{normQ1Trace}
\eea
The operators inside the trace in Eq. (\ref{normQ1Trace}) contain a macroscopic number of transfer matrices. Once again, this allows us to restrict to the Kronecker spaces which contain eigenvectors with eigenvalue 1. 
Let us introduce a shorthand notation for the derivatives of the reduced transfer matrices:\be
\mc T_1^{(n,n^\prime)}\equiv \left.\frac{\partial^n}{\partial u^n}\frac{\partial^{n^\prime}}{\partial\bar u^{n^\prime}}T_1(z=i,u,\bar u)\right|_{u=\bar u=1},
\ee
and likewise for $\mc T_2^{(n,n^\prime)}$. The derivatives in  Eq. (\ref{normQ1Trace}) yield\bea
\frac{\langle \mc I_1^\dagger \mc I_1\rangle}{N}&=&2\textrm{tr}\left\{(\mc T_1)^{N-1}\mc T_1^{(1,1)}+(\mc T_2)^{N-1}\mc T_2^{(1,1)}\right\}\nonumber\\
&&+2\sum_{n=0}^{N-2}\textrm{tr}\left\{\mc T_1^{(1,0)}(\mc T_1)^{n}\mc T_1^{(0,1)}(\mc T_1)^{N-2-n}\right\}\nonumber\\
&&+2\sum_{n=0}^{N-2}\textrm{tr}\left\{\mc T_2^{(1,0)}(\mc T_2)^{n}\mc T_2^{(0,1)}(\mc T_2)^{N-2-n}\right\}.\nonumber\\
\label{d1Ts}
\eea

In order for $\mc I_1$ to be quasilocal, the righthand side of Eq. (\ref{d1Ts}) must approach a finite value in the limit $N\to \infty$. First consider the last two terms in Eq. (\ref{d1Ts}). The derivatives of the $A_\alpha$ matrices at $u=1$ are\bea
A_0^\prime(1)&\equiv&\left.\frac{\partial A_0}{\partial u}\right|_{u=1}=i\sum_{r=0}^{m-1}\sin(\lambda r)|r\rangle \langle r|,\\
A_+^\prime(1)&\equiv&\left.\frac{\partial A_+}{\partial u}\right|_{u=1}=0,\\
A_-^\prime(1)&\equiv&\left.\frac{\partial A_-}{\partial u}\right|_{u=1}=-2\sum_{r=0}^{m-2}\cos(\lambda r)|r+1\rangle \langle r|.
\eea
Thus,\bea
\mc T_1^{(0,1)}&=&\mc T_1^{(1,0)}\nonumber\\
&=&\frac{i}2\sum_{r=0}^{m-1}\sin(2\lambda r)|r\rangle \langle r|+\frac{i}2\sum_{r=0}^{m-2}\sin(2\lambda r)|r+1\rangle \langle r|,\\
\mc T_2^{(0,1)}&=&i\sum_{r=0}^{m-1}\sin(\lambda r)\cos(\lambda r)|r\rangle \langle r|\nonumber\\
&&+i\sum_{r=0}^{m-2}\sin(\lambda r)\cos[\lambda (r+1)]|r+1\rangle \langle r|,\\
\mc T_2^{(1,0)}&=&i\sum_{r=0}^{m-1}\sin(\lambda r)\cos(\lambda r)|r\rangle \langle r|\nonumber\\
&&-i\sum_{r=0}^{m-2}\cos(\lambda r)\sin[\lambda (r+1)]|r+1\rangle \langle r|.
\eea
We then notice that \bea
\mc T_1^{(0,1)}|0\rangle&=&\mc T_1^{(1,0)}|0\rangle=0,\\
\mc T_2^{(0,1)}|0\rangle&=&\langle 0|\mc T_2^{(1,0)}=0.
\eea
These relations are a result of the decoupling of   state $|0\rangle$   from the other states at $u=1$ (see comment around Eq. (\ref{Izero})). Together with the projection in Eqs. (\ref{T1projector}) and (\ref{T2projector}), these relations imply that the last two terms in Eq. (\ref{d1Ts}) vanish in the thermodynamic limit.

We are   left with the contributions in the first line of Eq. (\ref{d1Ts}), which give\be
\lim_{N\to \infty}\frac{\langle\mc I_1^\dagger\mc I_1^{\phantom\dagger}\rangle}N=2\langle 0_L|\mc T_1^{(1,1)}|0\rangle+2\langle 0|\mc T_2^{(1,1)}|0\rangle.\label{normI1Ts}
\ee 
The derivatives of the reduced transfer matrices in Eq. (\ref{normI1Ts}) are\bea
\mc T_1^{(1,1)}&=&\sum_{r=0}^{m-1}\sin^2(\lambda r)|r\rangle \langle r| +2\sum_{r=0}^{m-2}\cos^2(\lambda r)|r+1\rangle \langle r|,\label{T111}\\
\mc T_2^{(1,1)}&=&\sum_{r=0}^{m-1}\sin^2(\lambda r)|r\rangle \langle r|+2\sum_{r=0}^{m-2}\cos(\lambda r)\cos[\lambda (r+1)]|r+1\rangle \langle r|.\label{T211}
\eea
The only nonzero matrix element that contributes to the norm of $\mc I_1$ is $\langle 0_L|\mc T_1^{(1,1)}|0\rangle.$ Using Eq. (\ref{zeroleft}), we obtain\be
\lim_{N\to \infty}\frac{\langle\mc I_1^\dagger\mc I_1^{\phantom\dagger}\rangle}N=4\left(1-\frac1m\right).\label{normI1}
\ee
This proves $\mc I_1$ is quasilocal for $m>1$.

Let us make some remarks about the conditions that lead to the norm's growing linearly with  system  size. This  is expected to happen whenever we have a representation that becomes  reducible for a specific value of a continuous parameter ($u=1$ in our case, see Eq. (\ref{Izero})) and the subrepresentation obtained in this case (the single state $|r=0\rangle$) is parity invariant. It is then clear   that   the two terms in Eq. (\ref{normfromTs}) are equal to $\Lambda^N$, where $\Lambda$ is the largest eigenvalue  in the parity-invariant subspace (assuming it dominates the norm). Expanding the operators around this specific value   with $\delta u=\varepsilon\ll1 $, we find that the transfer matrices    in Eq. (\ref{normfromTs}) behave as (here $\nu=1,2$)\be
T_\nu(1+\epsilon)=\left(\begin{array}{cc}
\Lambda&\varepsilon B_\nu\\
\varepsilon C_\nu&D+\varepsilon F_\nu
\end{array}\right),
\ee
where $D,F_\nu$ are   matrices in the subspace orthogonal to the parity invariant subspace. As a result, the eigenvalues behave as   $\textrm{Tr}\{T_{\nu}^N\}=(\Lambda+ \varepsilon^2 A_{\nu})^N\approx \Lambda^N+ N \varepsilon^2 A_{\nu}$. The norm will then be linear in $N$ as long as $A_1\neq A_2$.

\section{Nonlocal operators generated in the expansion of $\mc I(z,u)$\label{sec:nonlocal}}

The expansion in Eq. (\ref{expandIzu}) to higher orders in $\varepsilon =u-1$ gives rise to the family of operators\be
\mc I_\ell =\left.\frac{\partial^\ell }{\partial u^\ell}\mc I(z=i,u)\right|_{u=1}.
\ee
But $\mc I_1$ is the only quasilocal operator in this series because in general  the norm   of $\mc I_\ell$ scales like $N^\ell$ for $N\to \infty$. To see this, consider the case of $\mc I_2$:\bea
\langle \mc I^\dagger_2\mc I_2^{\phantom\dagger}\rangle &=&2\left.\textrm{ tr}\left\{\frac{\partial^4}{\partial u^2\partial \bar u^2}[T_1(z,u,\bar u)]^{N}\right\}\right|_{z=i, u=\bar u=1}\nonumber\\
&&-2\left.\textrm{tr}\left\{\frac{\partial^4}{\partial u^2\partial \bar u^2}[T_2(z,u,\bar u)]^{N}\right\}\right|_{z=i, u=\bar u=1}.\nonumber\\
&&\label{normQ2Trace}
\eea
When applying derivatives in Eq. (\ref{normQ2Trace}), we can discard terms which contain  $\mc T_1^{(0,1)}, \mc T_1^{(1,0)}, \mc T_2^{(0,1)}, \mc T_2^{(1,0)}$,  since their contribution vanishes in the thermodynamic limit. The result for large $N$ is \bea
\frac{\langle  \mc I^\dagger_2\mc I_2^{\phantom\dagger}\rangle}{N}&=&2\textrm{tr}\left[\mc T_1^{(2,2)}(\mc T_1)^{N-1}\right] +4\sum_{n=0}^{N-2}\textrm{tr}\left[\mc T_1^{(1,1)}(\mc T_1)^{n}\mc T_1^{(1,1)}(\mc T_1)^{N-2-n}\right]\nonumber\\
&& +2\sum_{n=0}^{N-2}\textrm{tr}\left[\mc T_1^{(2,0)}(\mc T_1)^{n}\mc T_1^{(0,2)}(\mc T_1)^{N-2-n}\right]-(\mc T_1\to\mc T_2).\label{d2Ts}
\eea
In contrast with Eq. (\ref{d1Ts}), the terms on the righthand side of Eq. (\ref{d2Ts}) that involve sums do not vanish identically because the matrices $\mc T_1^{(1,1)}, \mc T_1^{(2,0)}, \mc T_1^{(0,2)}$ do not annihilate the state $|0\rangle$ (and likewise for $\mc T_2$). In fact, the result of the sum increases  linearly with $N$ since the trace does not decay with the number  $n$ of $\mc T_1$'s between the derivatives. The coefficient of the $\mc O(N^2)$ term in the norm of $\mc I_2$ stems from terms in the sums with $n\sim N$. In order to extract this coefficient, we   insert another projection onto the eigenvectors of $\mc T_1$ or $\mc T_2$  with eigenvalue 1 and obtain\bea
\lim_{N\to\infty}\frac{\langle  \mc I^\dagger_2\mc I_2^{\phantom\dagger}\rangle}{N^2}&=&4\langle 0_L|\mc T_1^{(1,1)}|0\rangle^2-4\langle 0|\mc T_2^{(1,1)}|0\rangle^2+2\langle 0_L|\mc T_1^{(2,0)}|0\rangle\langle 0_L|\mc T_1^{(0,2)}|0\rangle\nonumber\\
&&-2\langle 0|\mc T_2^{(2,0)}|0\rangle\langle 0|\mc T_2^{(0,2)}|0\rangle.\label{Nslope}
\eea 
It is easy to verify that the last two terms in Eq. (\ref{Nslope}) cancel out. Using the matrices in Eq. (\ref{T111}) and (\ref{T211}), we find\be
\lim_{N\to\infty}\frac{\langle  \mc I^\dagger_2\mc I_2^{\phantom\dagger}\rangle}{N^2} =16\left(1-\frac1m\right)^2.
\ee

For general $\ell \geq 1$, the expression for the norm of $\mc I_\ell$ contains terms in which a number  $\ell $ of  matrices $\mc T_1^{(1,1)}$ are distributed over the $N$ sites of the chain. Since the contribution in the trace does not decrease with the separation between the $\mc T_1^{(1,1)}$'s, the norm grows with the number of ways to choose the positions of these matrices when they are far apart, therefore $\langle \mc I^\dagger_\ell \mc I_\ell^{\phantom\dagger}\rangle\propto N^\ell$. 

\section{Mazur bound\label{sec:mazur}}

Within linear response theory, the optical conductivity for a given model is related to the dynamical current-current correlation function via the Kubo formula. The real part of the optical conductivity can be written as\be
\sigma^\prime(\omega)=2\pi D\delta(\omega)+\sigma_{reg}(\omega),
\ee
where $D$ is the Drude weight and $\sigma_{reg}(\omega)$ is the regular part. A nonzero Drude weight implies infinite dc conductivity, {\it i.e.}, ballistic transport. The connection between integrability and   transport is made particularly clear by means of the Mazur bound \cite{zotos} for the Drude weight at finite temperature  $T$\be
D\geq\frac{1}{2LT}\sum_k\frac{|\langle J Q_k \rangle|^2}{\langle Q_k^\dagger Q_k^{\phantom\dagger}\rangle}.\label{mazurbound}
\ee
Here $L$ is the system size, $\langle\,\rangle$ denotes the thermal average, $J$ is the current operator and $\{Q_k\}$ is a set of operators that commute with the Hamiltonian and are orthogonalized in the form $\langle Q_k^\dagger Q_l^{\phantom\dagger}\rangle = \delta_{kl}\langle Q_k^\dagger Q_k^{\phantom\dagger}\rangle$. Although integrable models possess an infinite number of conserved quantities in the thermodynamic limit, it suffices to find one single operator that gives a nonzero contribution to the right hand side of Eq. (\ref{mazurbound}) in order to establish ballistic transport.   

The current operator is obtained from the continuity equation for the density of the conserved charge. The spin current operator for  the  XXZ model (\ref{XXZ}) reads \be
J=i\sum_j(\sigma_j^+\sigma_{j+1}^--\sigma_j^-\sigma_{j+1}^+).
\ee
This operator is odd under spin inversion:\be
\mc C^{-1}J\mc C=-J.
\ee
As   discussed in the  section \ref{sectionlocal},  all the local conserved quantities derived from the transfer matrix $t_{\mc Q}$ are invariant under spin inversion. This includes the XXZ Hamiltonian at zero magnetic field. As a result,  $\langle JQ_n\rangle=0$ for all the local $Q_n$'s. 

Let us now show that the quasilocal operator $\mc I_1$ in Eq. (\ref{defI1}) provides a nonzero Mazur bound at zero magnetic field.  Notice that, since $[\mc I_1,H]=0 $ exactly, there is no issue with the violation of the conservation law by boundary terms as in the open chain \cite{prosenbound}. We need to calculate the overlap between $J$ and $\mc I_1$\bea
\langle J \mc I_1\rangle &=&2^{-N}\textrm{Tr}\{J\mc I_1\}\nonumber\\
&=&Ni\sum_{\{\alpha_j\}}\frac{\partial}{\partial u}\textrm{tr}\left\{A_{\alpha_1}(u)\dots A_{\alpha_N}(u) -\left.A_{\alpha_1}^{t}(u)\dots A_{\alpha_N}^t(u)\right\}\right|_{u=1}\times\nonumber\\
&&\times2^{-N}\textrm{Tr}\{\prod_{j=1}^{N}\sigma_j^{\alpha_j}(\sigma_1^+\sigma_2^--\sigma_1^-\sigma_2^+)\}\nonumber\\
&=&\left.\frac{Ni}{2}\frac{\partial}{\partial u}\textrm{tr}\{[A_-(u),A_+(u)](A_0(u))^{N-2}\}\right|_{u=1}.
\eea
For $N\to \infty$, the factors of $(A_0)^{N}$ project the auxiliary space into  $|0\rangle$, which is the eigenvector of $A_0$ with eigenvalue 1. The nonzero contribution stems from applying to derivative to $A_-(u)$:\be
\langle J \mc I_1\rangle=-\frac{Ni}{2}\langle 0|A_+(1)A_-^\prime(1)|0\rangle=-N\sin \lambda.
\ee

\begin{figure}
\begin{center}
\includegraphics*[width=.5\columnwidth]{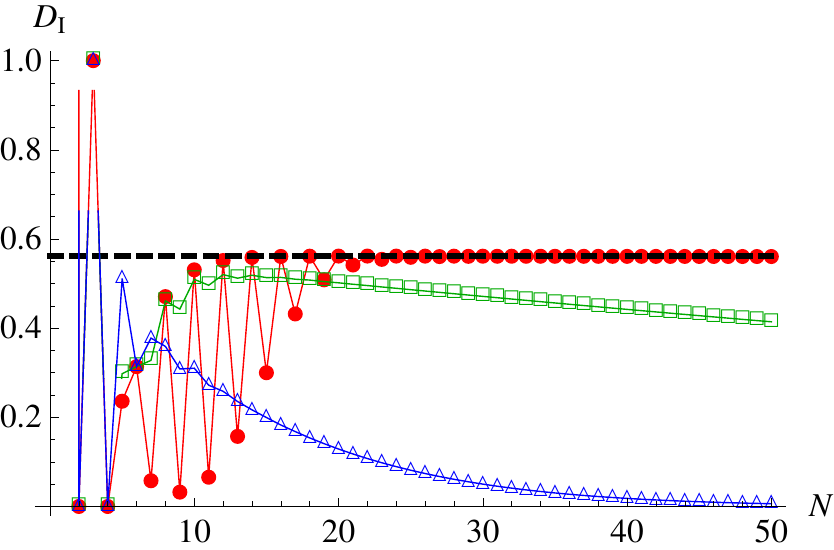}
\end{center}
\caption{Mazur bound from $\mc I(z=i,u)$ as a function of chain length $N$ for $q=e^{i\pi/3}$ and three different values of the parameter $u$: $u=1.01$ (red circles), $u=1.1$ (green squares) and $u=1.4$ (blue triangles). The dashed line represents the analytical result for $u\to 1$ and $N\to \infty$ in Eq. (\ref{Dzresult}).  \label{fig:Mazur}}
\end{figure}

Using the result for the norm  in Eq. (\ref{normI1}), we find that the contribution from $\mc I_1$ to the Mazur bound   in the high temperature limit is of the form $D\geq D_{\mc I_1}/4T$ with\be
D_{\mc I_1}=\lim_{N\to\infty }\frac{2\langle J\mc I_1 \rangle^2}{N\langle \mc I_1^\dagger \mc I_1^{\phantom\dagger}\rangle}=\frac{\sin^2\lambda}{2}\frac{m}{m-1}.\label{Dzresult}
\ee
This result agrees with the bound obtained from the quasi conserved operator for the open chain \cite{prosen1}.

Since $\mc I_0=\mc I(z=i,u=1)$ vanishes, the result in Eq. (\ref{Dzresult}) can also be written as\be
D_{\mc I_1}(z=i)=\lim_{N\to\infty}\lim_{u\to 1}\frac{2|\langle J \mc I(z=i,u)\rangle|^2}{N\langle \mc I^\dagger(z=i,u) \mc I(z=i,u)\rangle},\label{DZfinitesize}
\ee
where the order of the limits matters. 
The role of the limit $u\to 1$ before $N\to \infty$ is illustrated in Fig. \ref{fig:Mazur}, where we calculate the Mazur bound for finite chains numerically without using the projection into subspaces of largest eigenvalues.  For $u^4\neq 1$, the conserved  quantity $\mc I(z=i,u )$ is nonlocal and its contribution  to the Mazur bound decreases exponentially with system size. For small $|u-1|$ and large finite  $N$, the Mazur bound approaches a plateau that agrees with the analytical result in the thermodynamic limit. 

\section{Continuous family of quasilocal operators\label{sec:family}}
 
The choice of the spectral parameter $z=i$ in Eq.~(\ref{defI1}) is not required to derive a quasilocal  conserved quantity. Generalizing the results of section \ref{seccomputenorm} to arbitrary values of $z$, we find that the norm of \be\mc I_1(z)=\left.\frac{\partial}{\partial u}\mc I(z,u)\right|_{u=1}\ee
can be computed from the reduced transfer matrices ({\it cf.} Eqs. (\ref{defT1red}) and (\ref{defT2red}))
\bea
\mc T_1(z)&=&\sum_{r=0}^{m-1}\left\{\left[(\textrm{Im }z)^2\cos^2(\lambda r)+(\textrm{Re }z)^2\sin^2(\lambda r)\right]|r\rangle\langle r|\right.\nonumber\\
&&\left.+\frac{|z|^2}2\sin^2[\lambda (r+1)]|r\rangle\langle r+1|+\frac{|z|^{-2}}2\sin^2 (\lambda r)|r+1\rangle\langle r|\right\},\\
\mc T_2(z)&=&\sum_{r=0}^{m-1}\left\{\left[(\textrm{Im }z)^2\cos^2(\lambda r)+(\textrm{Re }z)^2\sin^2(\lambda r)\right]|r\rangle\langle r|\right.\nonumber\\
&&\left.-\frac12\sin(\lambda r)\sin[\lambda (r+1)]\left[|z|^2|r\rangle\langle r+1|+|z|^{-2}|r+1\rangle\langle r|\right]\right\}.
\eea

\begin{figure}
\begin{center}
\includegraphics*[width=.5\columnwidth]{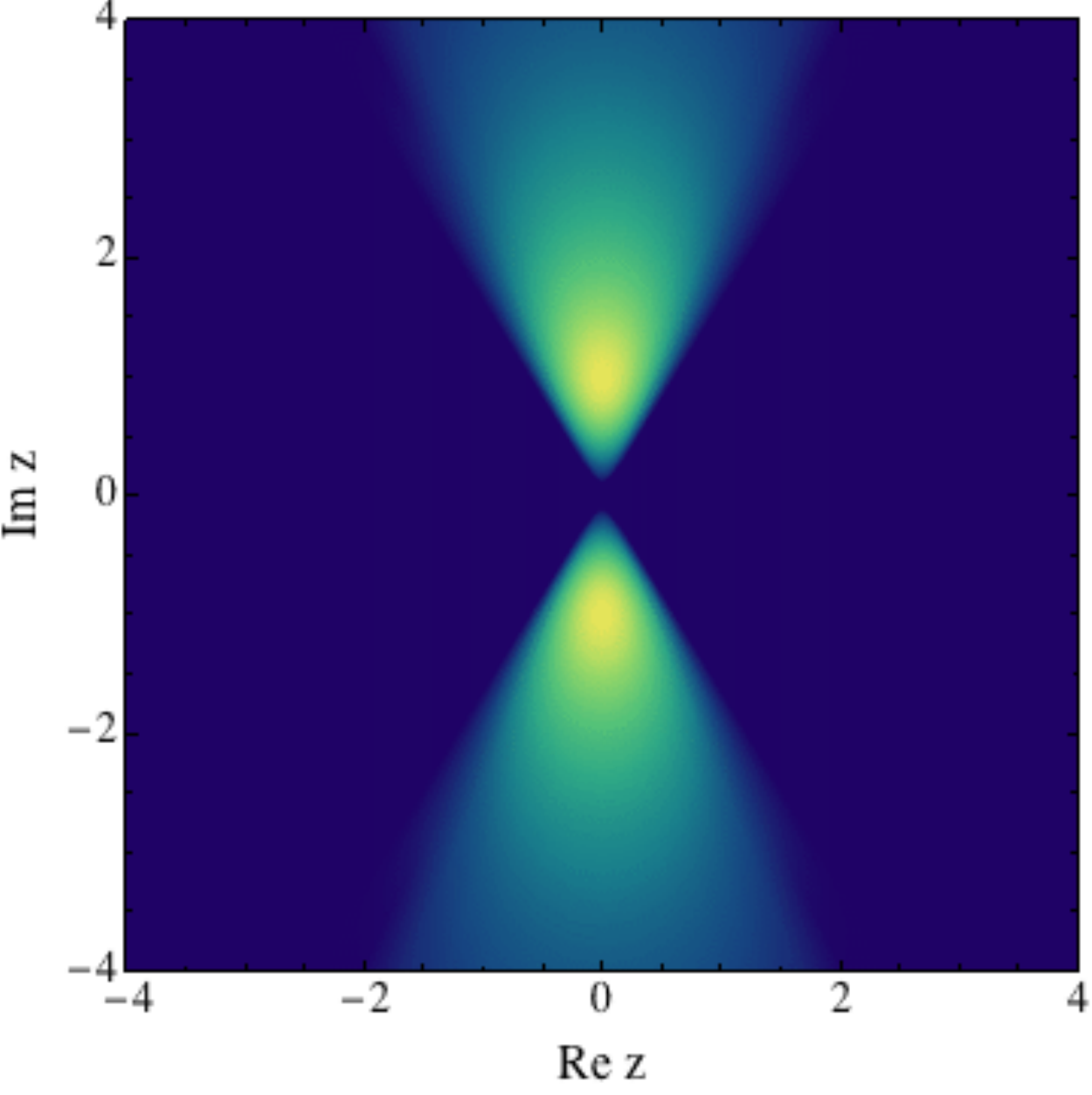}
\end{center}
\caption{Magnitude of the Mazur bound $D_{\mc I_1}(z)$ calculated using a single quasilocal operator $\mc I_1(z)$, as a function of the  spectral parameter $z$, for $q=e^{i\pi /3}$. Brighter regions correspond to larger values of $D_{\mc I_1}(z)$.  \label{fig:norm}}
\end{figure}

The state $|r=0\rangle$ is  an eigenvector of $\mc T_2(z)$ and a right  eigenvector of $\mc T_1(z)$, with eigenvalue $\Lambda=(\textrm{Im }z)^2$. As discussed at the end of section \ref{seccomputenorm}, the condition for quasilocality is that $\Lambda$ be the largest eigenvalue of both transfer matrices. As shown in \cite{prosenilievski}, this condition is satisfied by a continuous set of values of $z$ in the complex plane. 

As a measure of quasilocality, we use the Mazur bound in Eq. (\ref{DZfinitesize}) generalized to arbitrary $z$. A nonzero value of $D_{\mc I_1}(z)$ implies that the norm of $\mc I_1(z)$ is extensive. Fig. \ref{fig:norm} illustrates the magnitude of the Mazur bound $D_{\mc I_1}(z)$   in the complex $z$ plane for $q=e^{i\pi /3} $ ($\Delta=1/2$). We find that $D_{\mc I_1}(z)$ is maximum at $z=\pm i$, where it assumes the value predicted by Eq. (\ref{Dzresult}).  The domain where $D_{\mc I_1}(z)> 0$ was discussed in \cite{prosenilievski}: writing $z=|z|e^{i\theta}$, the conserved operator $\mc I_1(z)$ is quasilocal inside the cone $||\theta|-\frac{\pi}2|<\frac\pi{2m}$.

Therefore, the  Mazur bound obtained from a \emph{single} quasilocal operator is  maximized by the choice $z=\pm i$. However, the entire continuous  family  $\{\mc I_1(z)\}$ can be used to raise the bound. The idea is to replace the sum on the rhs of  Eq. (\ref{mazurbound}) by an integral over the spectral parameter $z$, using the orthogonality between different elements in the family, as done in  \cite{prosenilievski}.

\section{Conclusion\label{sec:concl}}

We have described a method to derive quasilocal operators which commute with the Hamiltonian of the XXZ chain with periodic boundary conditions. The key to this procedure is to introduce an auxiliary transfer matrix $t_{\mc A}(z,u)$ that depends on two parameters, namely the usual spectral parameter $z$ and the representation  parameter $u$. The latter is  a parameter of  the highest-weight representation for the quantum group algebra that arises in the Yang-Baxter relation for the Lax operator. For values of anisotropy $\Delta=\cos(\pi l/m)$, with $l,m$ integers, the highest-weight representation has  finite dimension $m$. The two-parameter conserved operator $\mc I(z,u)$ that has a nonzero overlap with  the spin current operator is defined  from a linear  combination of the auxiliary transfer matrix and its conjugate under parity. The norm of $\mc I(z,u)$ can be calculated using the transfer matrices $T_1(z,u)$ and $T_2(z,u)$, which are related by a similarity transformation. A quasilocal operator is obtained by expanding $\mc I(z,u)$ about the special value   $u=1$ where the highest-weight representation becomes reducible and the eigenvector of $T_1(z,u=1)$ and $T_2(z,u=1)$ with the largest eigenvalue decouples from the other states. The remaining spectral parameter $z$ labels a continuous family of quasilocal conserved quantities $\{\mc I_1(z)\}$. This is in contrast with the usual discrete set of local conserved quantities   which are   obtained by taking logarithmic derivatives of the transfer matrix $t_{\mc Q}(z)$ (defined with a spin-$1/2$ representation in   auxiliary space).

It has been shown that the quasilocal operators are important to set a nonzero lower bound for the Drude weight   in the spin-$1/2$ XXZ chain \cite{prosen1}. An important open question is whether there exist other families of quasilocal operators beyond the ones derived by this method. Additional conserved quantities may be expected from the observation that the Mazur bound computed from the set $\{\mc I_1(z)\}$ has a fractal $\Delta$ dependence \cite{prosen1,prosenbound} which is perhaps absent in the actual   Drude weight  at high temperatures \cite{karrasch}.

We note that, while here we have focused on the periodic chain, it should  be possible to apply the same techniques to integrable models with open boundaries, taking into account reflection operators at the boundaries.  In fact, in \cite{lazarescu} a two-parameter family of transfer matrices has been constructed for the open asymmetric simple exclusion process (ASEP) (see Eqs. (47) and (48) of \cite{lazarescu}, which are the generalization of the conserved quantities to the open case). The effect  of the boundary parameters on transport properties is an interesting open question.

The role of quasilocal conserved quantities  in the GGE also remains to be clarified \cite{wouters,mierzejewski,pozsgay}. Remarkably, there is evidence   that expectation values of local observables  in post-quench steady states deviate from the predictions of the GGE even for $\Delta>1$ \cite{wouters,pozsgay}, {\it i.e.} in the gapped N\'eel phase, where the method  described here does not yield any quasilocal operators.

\section*{Acknowledgements} 
We thank Fabian Essler, Andreas Kl\"umper, and Gr\'egoire Misguich for discussions. This work is supported by CNPq (R.G.P.), the SFB/TR 49 (J.S.), NSERC (J.S., I.A.),  and CIfAR (I.A.).

\appendix

\section{Similarity between reduced transfer matrices $\mc T_1$ and $\mc T_2$\label{appsimilar}}

The transfer matrices defined in section \ref{seccomputenorm} are
\bea
\mc{T}_1&=&\left(
\begin{array}{cccccc}
 1&  \frac12\sin^2\lambda & 0 &0&\dots &0\\
0& \cos^2\lambda & \frac12\sin^22\lambda&0&\dots&0 \\
0 &  \frac12\sin^2\lambda & \cos^2 2\lambda &  \frac12\sin^23\lambda&\dots&0\\
\vdots & \vdots & \vdots &\vdots&\ddots&\vdots\\
0 & 0 & 0 &0&\dots&\cos^2(m-1)\lambda\end{array}\label{matrixT1}
\right),\\
\mc{T}_2&=&\left(
\begin{array}{ccccc}
 1& 0 & 0 &\dots &0\\
0 & \cos^2\lambda &- \frac12\sin\lambda\sin2\lambda&\dots&0 \\
0 & - \frac12\sin\lambda\sin2\lambda & \cos^2 2\lambda &\dots&0\\
\vdots & \vdots & \vdots & \ddots&\vdots\\
0 & 0 & 0 & \dots&\cos^2(m-1)\lambda\end{array}
\right).\label{T2matrix}
\eea
First we note that the sign of the off-diagonal terms of $\mc T_2$ can be changed by applying the ``$Z_2$ gauge  transformation'' $\mc S|r\rangle = (-1)^r|r\rangle$. Defining $\tilde{\mc T}_2=\mc S^{-1}\mc T_2 \mc S$, we obtain \be
\tilde{\mc T}_2=\mathbbm 1-B^2+\frac12B\Delta B, 
\ee
which is to be compared to Eqs. (\ref{T1withdeltanadB}) and (\ref{T2withdeltanadB}).

Let us then show that $\mc T_1$ is similar to $\tilde{\mc T}_2$. It suffices to show that they have the same characteristic polynomial. The characteristic polynomial for $\mc T_1$ reads\be
\textrm{det}(\mc T_1 -x \mathbbm 1)=(1-x)\, \textrm{det}\left[(1-x) \mathbbm 1-\tilde B^2+\frac12  \tilde \Delta \tilde B^2 \right],
\ee
where $\tilde B$ is the diagonal   matrix $\tilde B=\sum_{r=1}^{m-1}\sin(r\lambda)|r\rangle \langle r|$ and $\tilde \Delta $ is the uniform hopping matrix $\tilde \Delta=\sum_{r=1}^{m-2}(|r\rangle\langle r+1|+|r+1\rangle\langle r|)$ in the $(m-1)$-dimensional space. Likewise, the characteristic polynomial for $\tilde{\mc T}_2$ (and also for $\mc T_2$, given the similarity between them) is \be
\textrm{det}(\tilde\mc T_2 -x \mathbbm 1)=(1-x)\, \textrm{det}\left[(1-x) \mathbbm 1-\tilde B^2+\frac12\tilde B\tilde \Delta \tilde B \right].
\ee
Since $\tilde B$ is invertible (all of its eigenvalues are nonzero for $\lambda=l\pi /m$ with $l,m$ coprimes), we can apply a similarity transformation inside the determinant sign as follows:\bea
\textrm{det}(\tilde\mc T_2 -x \mathbbm 1)&=&(1-x)\, \textrm{det}\left[\tilde B^{-1}\left((1-x) \mathbbm 1-\tilde B^2+\frac12\tilde B\tilde \Delta \tilde B \right)\tilde B\right]\nonumber\\
&=&(1-x)\, \textrm{det}\left[(1-x) \mathbbm 1-\tilde B^2+\frac12\tilde \Delta \tilde B^2 \right]\nonumber\\
&=&\textrm{det}( \mc T_1 -x \mathbbm 1).
\eea
This shows that $\mc T_1$ and $\mc T_2$ are similar.

\section{Left eigenvector of $\mc T_1$ with eigenvalue 1\label{appeigenv}}

Let $|0_L\rangle$ denote the left  eigenvectors of $\mc{T}_1$ with eigenvalue 1, which obeys \be
 \langle 0_L| \mc{T}_1 =\langle 0_L|.\label{zeroleftap}\ee
We expand $|0_L\rangle $ in the orthonormal basis of $\{|r\rangle\}$ vectors \be
|0_L\rangle=\sum_{r=0}^{m-1}v_r|r\rangle.\label{eigenT1}
\ee
Our problem is then to find the coefficients $v_r$. For short, we denote the matrix elements of $ \mc{T}_1$ as $\langle i| \mc{T}_1|j\rangle=t_{i,j}$. A useful identity is\be
t_{r,r+1}t_{r+1,r}=\frac14(1-t_{r,r})(1-t_{r+1,r+1}).\label{identityts}
\ee

The eigenvalue equation (\ref{zeroleftap}) is satisfied identically for the  column  $r=0$. This corresponds to the freedom of choosing the value of $v_0$ (or the normalization of $| 0_L\rangle$).  Let us turn to the next simplest equation,   the one stemming from the column $r=m-1$:\be
v_{m-2}t_{m-2,m-1}+v_{m-1}t_{m-1,m-1}=v_{m-1},
\ee
from which we get \be
v_{m-1}=\frac{t_{m-2,m-1}}{(1-t_{m-1,m-1})C_0}v_{m-2},
\ee
with $C_0=1$.  Next, the   equation for column $r=m-2$ reads  \be
v_{m-3}t_{m-3,m-2}+v_{m-2}t_{m-2,m-2}+v_{m-1}t_{m-1,m-2}=v_{m-2}.\label{v1iterate}
\ee
Using Eqs. (\ref{identityts}) and (\ref{v1iterate}), we obtain\be
v_{m-2}=\frac{t_{m-3,m-2}}{(1-t_{m-2,m-2})C_1}v_{m-3},
\ee
with $C_1=1-1/(4C_0)$.  In general, we find for $r=1,\dots,m-1$\be
v_{m-r}=\frac{t_{m-r-1,m-r}}{(1-t_{m-r,m-r})C_{r-1}}v_{m-r-1},\label{recurvr}
\ee
with $C_0=1$ and \be
C_r=1-\frac{1}{4C_{r-1}},\qquad 1\leq r\leq m-2.
\ee
This  relation express $C_r$ as a continued fraction and  the solution can be readily seen to be\be
C_r=\frac{r+2}{2r+2}.
\ee
In addition, we can use the explicit expression for the matrix elements of $\mc T_1$ in Eq. (\ref{matrixT1}), which gives\be
\frac{t_{r,r+1}}{1-t_{r,r}}=\frac{\frac12\sin^2\lambda r}{1-\cos^2\lambda r}=\frac12.
\ee
Thus Eq. (\ref{recurvr}) simplifies to \be
v_{m-r}=\frac{r}{r+1}v_{m-r-1},\quad r=2,\dots,m.
\ee
Writing   the coefficients $v_r$, $r=1,\dots,m-1$, in terms of $v_0$, we find\be
v_{r}=\left(1-\frac{r}{m}\right)v_0.
\ee
Finally, setting $v_0=1$ we obtain the  vector in Eq. (\ref{eigenT1})\be
|0_L\rangle=\sum_{r=0}^{m-1}\left(1-\frac{r}m\right)|r\rangle. 
\ee

\section{Calculating traces in the thermodynamic limit\label{appA}}

Let $M_1$ be an arbitrary $m\times m$ matrix (not necessarily Hermitean), where $m$ is the dimension of the auxiliary space $\mc A$. We want to compute tr$\{M_1\}$ using the  eigenvectors of $\mc{T}_1$. Since $\mc{T}_1$ is non-Hermitean, its right and left eignevectors are different:\bea
\mc{T}_1|R_j\rangle&=&\lambda_j|R_j\rangle,\\
\langle L_j|\mc{T}_1&=&\lambda_j\langle L_j|.
\eea
Nonetheless, the left and right eigenvalues are equivalent because $\mc T_1$ and $(\mc T_1)^t$ have the same characteristic polynomial \cite{marcus}. 

Right eigenvectors with different eigenvalues are not necessarily orthogonal, {\it i.e.} $\langle R_j|R_l\rangle\neq \delta_{i,j}$.  Let $\{|r\rangle\}$ denote the orthonormal basis  of vectors representing  sites in the auxiliary space. We can expand the vectors $|r\rangle$ in the non-orthogonal eigenvector basis in the form\bea
|r\rangle&=&\sum_jV_{r,j}|R_j\rangle,\\
|r\rangle&=&\sum_jW_{r,j}|L_j\rangle.\label{Wleft}
\eea
The transpose of Eq. (\ref{Wleft}) yields\be
\langle r|=\sum_j\langle L_j| W_{r,j}=\sum_j\langle L_j| W^t_{j,r}.
\ee
The inverse transformation reads\bea
|R_i\rangle&=&\sum_r(V^{-1})_{i,r}|r\rangle,\\
|L_i\rangle&=&\sum_j(W^{-1})_{i,r}|r\rangle.
\eea
It can   be proved that the left eigenvectors are orthogonal to right eigenvectors with different eigenvalues (and degenerate eigenvectors can be orthogonalized) \cite{marcus}, so that\be
(W^{-1})^t V^{-1}=D,
\ee
with \be
D_{j,l}=\langle L_j|R_l\rangle=d_j\delta_{j,l}.
\ee
Thus the inverse of $D$ is also diagonal:\be
(W^t V)_{j,l}=\frac1{d_{j}}\delta_{j,l}.
\ee

The trace of $M_1$ can be written as\bea
\textrm{Tr}\{M_1\}&=&\sum_r\langle r|M_1|r\rangle\nonumber\\
&=&\sum_{r,j,l}W^t_{j,r}V_{r,l}\langle L_j|M_1 |R_l\rangle\nonumber\\
&=&\sum_{j,l}(W^tV)_{j,l}\langle L_j|M_1 |R_l\rangle\nonumber\\
&=&\sum_{j}\frac{\langle L_j|M_1 |R_j\rangle}{d_j}.
\eea
In our case $M_1$ is a product of a large number ($\sim \mc O(N)$) of transfer matrices  $\mc{T}_1$ on the left and on the right. In the thermodynamic limit the trace is dominated by the contributions from the eigenvector of $\mc{T}_1$ with eigenvalue $\lambda_j=1$:\be
|R_1\rangle = |0\rangle,\qquad \langle L_1|=\langle0_L|.
\ee
Thus
\be
\lim_{N\to\infty }\textrm{Tr}\{M_1\}= \frac{\langle L_1|M_1 |R_1\rangle}{d_1}=\frac{\langle 0_L|\tilde M_1 |0\rangle}{\langle0_L|0\rangle},
\ee
where $\tilde M_1$ is obtained from $M_1$ by dropping the factors of $\mc T_1^N$. Using eigenvectors normalized as in Eqs. (\ref{zeroleft}),  we can write simply
\be
\lim_{N\to\infty }\textrm{Tr}\{M_1\}=\langle 0_L|\tilde M_1 |0\rangle.\label{calctrace}
\ee

The following relation is also useful:\bea
\sum_r |r\rangle\langle r|&=&\sum_{r,j,l}V_{r,j}W^t_{l,r} |R_j\rangle\langle L_l|\nonumber\\
&=&\sum_{j,l}(W^tV)_{l,j} |R_j\rangle\langle L_l|\nonumber\\
&=&\sum_{j}\frac{ |R_j\rangle\langle L_j|}{d_j}.
\eea
If there is a large number of  $\mc T_1$'s on both sides we can project onto the eigenvectors with eigenvalue $\lambda_j=1$\be
\lim_{n\to\infty}\left(\mc{T}_1^n\sum_r |r\rangle\langle r|\mc{T}_1^n\right)=\frac{ |0\rangle\langle 0_L|}{\langle0_L|0\rangle}=|0\rangle\langle 0_L|.\label{closureT1}
\ee

We also need to calculate traces involving $\mc T_2$. These are easier because $\mc T_2$ is symmetric and the eigenvector with eigenvalue 1 is simply $|0\rangle$. Thus the trace of a matrix with $\mc O(N)$ factors of $\mc T_2$ can be reduced to\be
\lim_{N\to\infty}\textrm{Tr}\{M_2\}=\langle 0|\tilde M_2 |0\rangle .\label{calctrace2}
\ee
The equivalent of Eq. (\ref{closureT1}) for $\mc{T}_2$ is\be
\lim_{n\to\infty}\left(\mc{T}_2^n\sum_r |r\rangle\langle r|\mc{T}_2^n\right)=|0\rangle\langle 0|.\label{closureT2}
\ee

\end{document}